\documentstyle[12pt,epsf,aasms4,flushrt]{article}

\newcommand{\etal}{{\it et al.~\ }}
\font\email = cmtt12
\def\Msun{\ifmmode M_{\odot} \else $M_{\odot}$\fi}
\def\Lsun{\ifmmode L_{\odot} \else $L_{\odot}$\fi}

\def\mic{{$\mu$m}}

\begin{document}

\title{AN EMPIRICALLY BASED MODEL FOR PREDICTING INFRARED LUMINOSITY FUNCTIONS,
DEEP INFRARED GALAXY COUNTS AND THE DIFFUSE INFRARED BACKGROUND}

\author{M. A. Malkan}
\affil{Physics and Astronomy Department, University of California Los Angeles,
Los Angeles, CA 90095-1562}
\centerline{\email malkan@astro.ucla.edu}

\author{F. W. Stecker}
\affil{Laboratory for High Energy Astrophysics, NASA Goddard Space Flight 
Center, Greenbelt, MD 20771}
\centerline{\email stecker@lheapop.gsfc.nasa.gov}

\begin{abstract}

We predict luminosity functions and number counts for 
extragalactic infrared sources at various wavelengths
using the framework of our empirically based
model. Comparisons of our galaxy count results with 
existing data indicate that either galaxy 
luminosity evolution is not much stronger than $Q = 3.1$, where $L \propto (1+z)^Q$,
or that this evolution does not continue beyond a redshift of 2.
However, a derivation of the far infrared background from {\it COBE-DIRBE} 
(Cosmic Background Explorer--Diffuse Infrared Background Experiment) data 
suggests a stronger evolution for the far-infrared emission,
with $Q > 4$ in the redshift range between 0 and 1.
We discuss several interpretations of these results, and also discuss
how future observations can reconcile this apparent conflict.
We also make predictions of the redshift distributions of extragalactic
infrared sources at
selected flux levels which can be tested by planned detectors.
Finally we predict the fluxes at which various future surveys
will become confusion limited.

\end{abstract}


\section{Introduction: Motivation for Extending Previous Work}

Observational cosmology is just starting to benefit 
from unprecedented sensitivity gains
at wavelengths longward of 1 $\mu$m.
A prime example is the breakthrough achieved by the COBE satellite in the
first detections of the Cosmic Infrared Background (CIB) radiation
at wavelengths from 2 to 300 $\mu$m.
These measurements provide crucial information about the integrated
luminosity of galaxies.  Especially at the longer wavelengths, where
the K-corrections of individual galaxies actually reverse sign, 
the CIB measurements constrain the evolution of the
galactic luminosity function (LF) back to cosmologically interesting
redshifts ($z \ge $0.5).  

To elucidate these constraints, we started with the simple 
empirically based model which we used previously to calculate the CIB
(Malkan and Stecker 1998, MS98 hereafter). In MS98, we made use of
infrared observations of galaxies over a wide range of luminosities
and at various wavelengths.
We assumed that the systematic dependence of galaxy spectra with
luminosity which we observe today also applies at earlier
cosmic times.  We then started with the present-day infrared luminosity
function of galaxies, and assumed a pure luminosity evolution for
each galaxy, with $L \propto (1+z)^{3.1}$ out to a redshift of $z_{flat}$
of 1 or 2, beyond which no further evolution occurred.  

In this paper, we use the framework of  our 
empirically based model (MS98) to derive luminosity
functions at various infrared wavelengths and to make infrared galaxy 
count predictions. These calculations should provide timely predictions which
will aid in the planning and interpretation of ongoing and new deep infrared 
imaging observations and their implications in understanding  galaxy 
formation and evolution. 
We also re-examine our CIB calculations and compare them with the recent data.

Several new telescopes have, or soon will obtain
deep extragalactic number counts of infrared and millimeter sources
in large regions of the sky at high galactic latitude which are relatively 
free of galactic foreground infrared emission.
Surveys have recently been completed with the {\it ISOCAM} Infrared
Space Observatory Camera) and {\it ISOPHOT} (Infrared Space Observatory 
Photometer) imagers
on the Infrared Space Observatory, {\it ISO}
and with {\it SCUBA} (Submillimeter Common User Bolometer Array) 
on the James Clerk Maxwell 
Telescope. Several new ones are planned, {\it e.g.}, with {\it SIRTF} ( 
Space Infrared Telescope Facility), 
{\it SOFIA} (Stratospheric Observatory for 
Infrared Astronomy), {\it IRIS}
(Infrared Imaging Surveyor), 
and {\it FIRST} (Far Infrared and Submillimeter Telescope).

Even though spectroscopic redshifts are difficult to obtain for more
than a handful of the faint infrared sources, there is a high likelihood that
a significant fraction of them is at cosmologically interesting
distances. Thus, deep galaxy number counts of sufficiently large
areas will provide statistical
information about galaxies at large lookback times, and therefore about
galaxy formation and evolution.
We use our model here to predict such number counts.
We also re-examine our predictions of the spectral 
energy distribution (SED) of the CIB,
in light of the subsequent {\it COBE} results.

If redshifts can also be measured (spectroscopically or photometrically) 
for statistically significant subsamples of faint
infrared sources, even more details of the evolution can be determined.
Therefore, in this paper we also make predictions of the redshift 
distributions of galaxies at
selected flux levels which will be accessible in the future.
Finally, we use the model predictions to estimate the flux levels
at which various surveys should become confusion limited.

\section{Galaxy Spectral Energy Distributions}

The two key factors determining the infrared counts 
are the evolution of star formation rates,
and the amount and distribution of dust grains
in galaxies as a function of luminosity and time.
Growing observational evidence indicates that the dust contents of
high-redshift galaxies was not very different from what is observed in
present day galaxies (see Malkan 1998; 2000 for reviews).
The more luminous low-redshift galaxies contain enough dust around
their actively star-forming complexes to absorb much of the ultraviolet
continuum light, and re-emit it in the mid infrared and far infrared.
The ratio of infrared luminosity to blue and ultraviolet luminosity
can exceed one, and this ratio is observed to be a {\it systematically
increasing function of galaxy luminosity}, or current star formation rate
(Spinoglio {\it et al.} 1995).
Furthermore, the enhanced dust emission in luminous galaxies arises
from relatively warm grains which emit at 25 to 60 $\mu$m.
We have made the simplifying assumption that the same systematic
correlations of dust content and resulting spectral shape as a function
of luminosity that we measure at $z\sim 0$ also apply to galaxies
at larger redshifts. 

MS98 used the correlations of bolometric luminosity with the
luminosities at 12, 25, 60 and 100 $\mu$m
from Spinoglio \etal (1995) to define the spectral shape of
galaxies as a function of their luminosities at $L_{60}$ or $L_{12}$,
their luminosities at 60 $\mu$m and 12 $\mu$m respectively.
In the near infrared, we assume luminosity independent spectral shapes 
with intrinsic zero-redshift average colors of:
$V-K = $[0.55$\mu$m--2.2$\mu$m] = 3.2 mag, $I-K = $[0.8$\mu$m--2.2$\mu$m] = 2.0 mag,
$J-H = $[1.2$\mu$m--1.6$\mu$m] = 0.9 mag and  $H-K = $[1.6$\mu$m--2.2$\mu$m] = 0.4 mag. 

At long wavelengths, MS98 assumed greybody thermal emission appropriate
for dust grains with emissivity $\propto\lambda^{-1.5}$,
following the correlation in Appendix B of Spingolio \etal (1995).
In this paper we have improved the description of the emission longward of
100$ \mu$m, the last {\it IRAS} (Infrared Astronomy Satellite) band.  
We have used the growing sets of 60 to 200 
$\mu$m galaxy photometry data becoming available from the {\it ISOPHOT} 
instrument on {\it ISO}.  As summarized by Spinoglio {\it et al.}
(2000), new far-infrared photometry is now available for dozens of
nearby galaxies which span a wide range in luminosity.  These observations
confirm the prediction of Spinoglio \etal (1995) that the galaxies with
``warmer" (i.e. bluer or flatter) 60 to 100$ \mu$m spectral slopes also 
appear warmer in
the 100 to 200 $\mu$m region, where their thermal dust spectra
approach a Rayleigh-Jeans distribution modified by  
dust emissivity $\propto \lambda^{-n}$ where $1.5 \le n \le 2$. We also 
confirm their finding that the
warmer dust (associated with regions of recent star formation) increases
the 60 $\mu$m flux relative to the flux longward of 100 $\mu$m systetmatically 
in the more infrared-luminous galaxies.  

Thus we are able to use these 
{\it ISOPHOT} observations to make a direct calibration of the average 
long-wavelength spectrum of galaxies as a function of their 60 $\mu$m 
luminosity.  We describe the 100 to 200$\mu$m spectra as a broken power law
with a break at 145 $\mu$m.  The slopes above and below 145$\mu$m
have the same luminosity dependence
$$\alpha_{100-145} = -0.65 + 0.6 Log (L / L_*) \eqno{(1a)} $$
and
$$\alpha_{145-200} = +2.0 + 0.6 Log (L / L_*) \eqno{(1b)} $$
where $L_*$ is the luminosity of a typical normal galaxy at 60$\mu$m,
$10^{43}$ erg s$^{-1}$.

At a given luminosity, the overall shapes of our average infrared spectra 
(from MS98) agree well with
those used in other published calculations (Pearson \etal 2000;
Xu \etal 2000; Takeuchi \etal 2000).  There is a tendency for our
models to predict more flux around 5 $\mu$m in luminous starburst
galaxies.  We believe that our SEDs in this wavelength range are more 
realistic because they are empirically 
based on interpolations using 3.5 and 12 $\mu$m photometric databases of
a large number of galaxies. Therefore, they take 
account of the very hot dust which is quite often found to produce
an ``excess" 3.5 $\mu$m flux, over and above that obtained by merely 
extrapolating stellar photospheric fluxes longward of 2 \mic.

\section{Galaxy Luminosity Functions}

In our calculation, the four luminosity relations obtained by 
Spinoglio {\it et al.} (1995) at 12, 25, 60 and 100 $\mu$m\
(and our estimates at 2.2 and 3.5 $\mu$m)
were inverted so that a luminosity at any given rest wavelength could
be determined from
the 60 $\mu$m luminosity, $L_{60}$.
This allowed us to 
make a mapping of the 60 $\mu$m luminosity function  (LF) of Lawrence \etal (1986)
into LFs at any infrared wavelength using the transformation relation
$$\phi_{\lambda}(\log L_{\lambda}) = \phi_{60}(\log L_{60}) (d\log L_{60} / 
d\log L_{\lambda})\eqno{(2)}$$

Given the fairly good scaling of other wavelengths with the 60 $\mu$m
luminosity, the Jacobian term on the right (which conserves total
number of galaxies) varies from 1.0 in the  25 to 60 $\mu$m range, up to 1.25
at 12 and 300 $\mu$m.

MS98 made calculations with 60 $\mu$m LFs from Saunders {\it et al.} (1990) and
from Lawrence {\it et al.} (1986).
In this paper we adopt the parameters from Lawrence \etal (1986), {\it viz.},
$\alpha=1.7$ and $\beta=1.8$ with a normalization constant of $C = 4.07 \times 10^{-4}$,
because they give a better match to the local 60 $\mu$m counts.
In Figure 1 we show the galaxy luminosity functions at a redshift of
0 based on our model.  

The 12$\mu$m LF has essentially the identical shape to the {\it bolometric}
LF of non-Seyfert galaxies, because the 12$\mu$m flux is a constant
6\% of the bolometric flux for all galaxy types 
except normal ellipticals (Spinoglio \etal 1995).
By comparison, the 25 $\mu$m and 60$\mu$m LFs are flatter.  They extend out
more strongly to high luminosities because high-L galaxies emit a 
relatively larger fraction of their bolometric luminosities at
25 to 60$\mu$m.  Conversely, the 3 $\mu$m and 400$\mu$m LFs are
much steeper.  This is because it is the less luminous galaxies which
emit relatively more power at 3 $\mu$m and 400$\mu$m.

For comparison, we also plot estimated local luminosity functions
at 12, 25, 60 and 100$\mu$m, taken from the literature.
The 25 $\mu$m and 100$\mu$m data ($\times$'s and open squares, respectively)
are from Shupe \etal (1998) and Soifer and Neugebauer (1991). 
The particular fiducial LF we have used was specifically fitted to the
60$\mu$m data points (shown as solid squares). 
Thus it is not surprising that the model LF (solid line)
matches these observations extremely well.
The validity of our wavelength transformation equations is 
confirmed by comparison of our model LF's with data at longer
and shorter wavelengths.
Our model LF at 12$\mu$m (dot-dash line) agrees well with the
data from Rush, Malkan \& Spinoglio (1993, solid circles, 
which include AGN in the totals). 
New determinations of the 12$\mu$m and 15$\mu$m LF's
(open triangles from Xu \etal 1998), and at 25 $\mu$m
find substantially fewer
low luminosity galaxies.  One reason for this is understood:
in contrast to the other LF estimates, these two new determinations
were corrected for the overdensity of nearby low luminosity galaxies
in the Virgo cluster.  These two corrected LFs thus may be
be more representative of the present day Universe.
However, if we are using LF's to predict number counts of bright
sources in the Northern sky, the actual local LF of Rush {\it et al.} (1993)
would give more accurate results. On the other hand,
our 100 $\mu$m model LF has a slightly flatter slope than the data, 
but with essentially identical overall normalizations.
Our use of the Lawrence 
{\it et al.} (1986) LF is an adequate compromise
fit to all of the local {\it IRAS} data from 12 to 100$\mu$m.

Fortunately, none of these subtleties makes a significant difference
in the predicted counts.  This is because the characteristic bent 
shape of the LF guarantees that the counts at any flux level are
always dominated by the number of galaxies around the characteristic
position of the knee in the LF, defined to be $L_*$.
As long as all the LF's are accurate around $L_*$ (where there is
virtually no disagreement), they will yield almost exactly the
correct source number counts.
We confirmed this by recalculating the predicted CIB for the Saunders {\it et al.} 
LF. Holding all other parameters constant, 
it is about 10\% lower at essentially all infrared wavelengths 
than the CIB prediction based on the Lawrence {\it et al.} LF.

Since MS98 demonstrated that active galaxies (AGN) contribute less than
10\% of the diffuse extragalactic infrared background, a fraction within
the errors of the models and the data, we neglect the AGN component to
the CIB in this work.  Xu \etal (2000) also reached the same conclusion.

\section{Galaxy Evolution}

The luminosity functions at higher redshifts were calculated using the
pure luminosity evolution relations

$$\phi[L_{\nu}, \lambda; z] = \phi_{MS,z=0}
[L_{\nu /(1+z)}/(l+z)^Q,(1+z)\lambda] 
\eqno{(3)}$$
where $$\phi_{MS, z=0} = 
[C (L/L_{*})^{1-\alpha}[1 + (L/\beta L_{*})]^{-\beta}]_{60}
d\log\phi_{60}/d\log\phi(\lambda)\eqno{(4)}$$
where the subscript 60 refers to a wavelength of 60 $\mu$m and 
$Q$ is the galaxy luminosity evolution parameter,
respectively. 
We note that these are differential luminosity functions and that we follow
the common convention where they are measured per unit logarithmic interval
of luminosity.

We adopt here the same evolutionary assumptions as MS98, {\it viz.},
pure luminosity evolution with all galaxy luminosities
scaling as $(1+z)^Q$ up to a redshift $z_{flat}$ and no evolution for
$z_{flat} \le z \le z_{max}$ with a cutoff at $z_{max} = 4$. 

As observations of high-redshift galaxies improved, they have tended
to reinforce our earlier redshift evolution assumptions as providing
a realistic description of the cosmic evolution of galaxy luminosities 
(Madau, Pozzetti \& Dickinson 1998, Steidel, \etal 1999, Blain \& Natarajan 
2000; Hopkins, Connolly \& Szalay 2000).
More complex evolutionary histories could be imagined.
However, our simple two-parameter formulation captures most of the
range of significant possibilities, and the observational data are
far too limited to try to constrain the values of a third possible
parameter.

The ``Best Estimate" galaxy evolution model used by MS98 to predict the
diffuse infrared background took $Q = 3.1$ and $z_{flat}= 2$.  
\footnote{In generating Figure 2 of MS98, 
galaxies above $z = 2$ were erroneously assumed to have the same
luminosities as at $ z = 0$. 
Here we more properly assume that the LF has evolution up to   
$z_{flat}$ and no  further evolution beyond that. Our corrected result and
that given in MS98 are not significantly different.
We have also corrected an error in the short-wavelength extrapolation of
the Spinoglio \etal (1995) SEDs. As discussed in Section 2, we now use average
optical colors of late-type spirals.}
Here we refer to this model as the ``Baseline" model.
             
MS98 also showed a more conservative calculation, in which the
luminosity evolution (with $Q$ still equal to 3.1) was frozen at
$z_{flat} \ge$ 1.  Thus by $z = 2$, this conservative scenario assumes 
galaxy luminosities
were only $2^{3.1}$ times more luminous than today, rather than the
factor of $3^{3.1}$ assumed in the baseline model.
In this paper, we refer to the $z_{flat}$ = 1; $Q = 3.1$ model as the
``Lower Limit" scenario.

We also consider here an alternate ``Fast Evolution" scenario, in which 
the evolution index is increased to $Q = 4.1$, while
$z_{flat}$ is taken to be 1.3.
As for the baseline case, this scenario also implies that galaxies were
forming stars at $\sim$30 times higher rates at $z = 2$ than today.
However, with fast evolution, the galaxy luminosities at $z = 1.3$ were
also $\sim$30 times higher, in contrast to the baseline model in which they 
were 8.6 times their current luminosities. Support for the Fast Evolution 
scenario is found in the recent {\it NICMOS} (Near Infrared Camera and 
Multi-Object Spectrograph) studies of Hopkins \etal (2000).

\section{Diffuse Infrared Background Radiation}

First we compare the predictions of our model with the
new measurements of the CIB as shown in Figure 2.
The baseline model with $Q = 3.1$,  $z_{max}$ = 4 and $z_{flat}$ = 2,
is shown by the middle solid  line. The lower dashed curve is the prediction
of our lower limit model with $z_{flat}$ = 1.
The upper (dot-dashed) curve shows our fast evolution ($Q = 4.1$; 
$z_{flat}$ = 1.3) case.

The predicted CIB is most sensitive to the exponent of 
the luminosity evolution, $Q $.  Increasing or decreasing $Q$
by 0.5 results in a CIB which increases or decreases by 25 to 30\%
at most wavelengths. The second most important parameter is the
redshift at which the luminosity evolution stops, which was
$z_{flat} = 2$ in the baseline calculation.  If $z_{flat}$ is dropped to
1 with Q unchanged, the CIB drops by about 25\% at most wavelengths, as is evident from the lower dashed curve.

If we extend the evolution in our models to values of
$z_{flat}$ significantly above 2, they would predict too much flux in the FIR-sub mm
range to be consistent with the
FIRAS results (Fixsen \etal 1998). 

Finally, the least important free parameter is the maximum 
redshift for which infrared emitting galaxies are included, $ z_{max} = 4$
in the baseline case.  When this $z_{max}$ is decreased to 2, the CIB
decreases by about 30\% at most wavelengths. The combination of 
redshift evolution and cosmology insures that it is the galaxies at redshifts
$\sim$1 which contribute most to the CIB, as discussed further below. 

In our lower limit case, the redshift of maximum evolution is
reduced to $z_{flat}=1$ while $Q$ is taken to be to 3.1, 
resulting in a CIB prediction about 70\% lower than our baseline case.
Our fast evolution case gives CIB fluxes which are about 30\% higher than our
baseline case at mid infrared and far infrared wavelengths.
All of our models are consistent with the subsequent 
{\it COBE} detections of the cosmic
background at 2.2 $\mu$m and 3.5 $\mu$m (Dwek and Arendt 1998;
Gorjian, Wright \& Chary 2000). 

We note that the best constraint on the CIB at mid-infrared wavelengths comes
from studies of the lack of absorption features in blazar spectra up to 10 TeV.
This gives an upper limit on the 20 $\mu$m flux of 4 to 5 nW m$^{-2}$sr$^{-1}$
(Stecker \& de Jager 1997, Stanev \& Franceschini 1998; Biller 
\etal 1998). This limit, when combined with our predicted CIB SEDs,
disfavors evolution with $Q > 5$. 

The {\it COBE-DIRBE} far infrared determinations seem to favor 
strong evolution, at least to a redshift $\sim$1,   
{\it i.e.}, our fast evolution case which is the upper curve in Figure 2,
generated using $Q = 4.1$ and $z_{flat} = 1.3$.
This case is within 2 $\sigma$ of the results of the
Hauser \etal (1998) analysis of the {\it COBE-DIRBE} data at 140 and 240
$\mu$m. When one considers that these results were themselves at the 4 $\sigma$
level and that there still may have been some 
undersubtraction of foreground emission, we consider our fast evolution case to
be consistent with both the {\it COBE-DIRBE} and {\it FIRAS} numbers.

We also show points at 140 and 240 $\mu$m derived from the {\it COBE-DIRBE} 
data using the {\it FIRAS} calibration, which suffered from smaller
systematic errors than the {\it COBE-DIRBE} calibration (Fixsen \etal 1997).
This calibration lowers the flux numbers and brings them into better agreement
with our fast-evolution and baseline model calculations.

The point at 100 $\mu$m was identified by Lagache \etal (2000) as the 
extragalactic background flux. However, it may actually be an upper limit, 
since their analysis did not demonstrate isotropy (Hauser \& Dwek 
2001).\footnote{There was tentative detection of the CIB at 60 and 
100 $\mu$m by Finkbeiner, Davis \& Schlegel (2000).
However, this result now appears to have suffered from contamination by local
dust emission (Finkbeiner 2001) and is not shown in Figure 2 for this reason.}

\section{Observational Appearance of a Typical Evolving Galaxy}

The redshift distribution can be understood by considering the
brightness of a typical bright galaxy, viewed out to various redshifts.
For example, let us consider a galaxy with a current 60$\mu$m luminosity of
$L_{*}$ at the knee in the LF.
The observed flux of a galaxy whose present luminosity is $L_{*}$, given as a 
function of redshift, is shown in Figure 3 for observing wavelengths of 25, 
60 and 200$\mu$m. Observing an $L_{*}$ galaxy at a fixed 
wavelength samples emission at progressively shorter rest-system
wavelengths at progressively higher redshifts.
For a wavelength of 200$\mu$m, the peak of the far infrared spectrum of this 
galaxy is shifted into the observing bandpass at $z \sim 1$, resulting
in a negative K-correction.  This explains the relatively
flat portion of the flux curve from $z = 0.6$ to 1.0.
At higher redshifts, luminosity evolution produces an even
stronger effect, which causes all the flux curves to become
relatively flat from $z = 1.5$ to 2.0. 

Figure 3 also shows the 60$\mu$m fluxes of galaxies which are currently
100 times more or less luminous than $L_{*}$, indicated by the
short and long dashed lines respectively. At an observing wavelength 
of 60$\mu$m, evolutionary changes in galaxy spectra
make the less luminous galaxies relatively
brighter at high redshifts. 

\section{Galaxy Number Counts}

We can sum our generated family of LFs over redshift to obtain the
expected number of observable sources at a given flux level and wavelength
over the entire sky:

$$dN/dF(\lambda_{0}) = 16 \pi (c/H_{0})^3  \int \displaylimits_{0}^{z_{max}}
 dz \int \displaylimits_{L_{min}(F,z)}^{\infty} (d\log L) \phi(\lambda_{0}/(1+z),z)
(1+z)^{-5/2} (\sqrt(1+z) - 1)^2 \eqno{(5)}$$
assuming $\Omega = 1$ and $\Lambda = 0$.
Figure 4 shows the predicted integrated galaxy counts as a function of 
minimum observable
flux at each of the 8 logarithmically spaced wavelengths
shown in Figure 1. 

\subsection{Results}

As expected, the slopes of the integrated counts are approximately
equal to the Euclidean value (-3/2) at the brighter flux levels (above 1 mJy).
The galaxy counts from 2 $\mu$m to 100 $\mu$m
have similar slopes with a Euclidean (high-flux)
normalization (in Jansky flux units) 
given by the formula:
$$Log F^{3/2}N(>F) = 0.46 - 0.06 Log (\lambda) +0.83 (Log \lambda)^2\eqno{(6)}$$
reflecting the average spectral spectral energy distribution of a typical
$L_{*}$ galaxy. However, the characteristic knee where the slope
flattens is shifted to higher flux levels at longer wavelengths.
This reflects the fact that an $L_{*}$ galaxy is more luminous
at longer wavelengths, and can therefore be seen to larger distances.
The 3 and 6$\mu$m counts rise above the Euclidean value briefly 
around 0.1 mJy, owing to the positive K-corrections at those wavelengths.
At fluxes fainter than 0.1 mJy, the counts at all wavelengths are 
flatter than the Euclidean slope, because the deviations
due to cosmology dominate over galaxy evolution,
which was assumed to level off at $z > 2$.
At 50 and 100$\mu$m, this cosmological flattening of the counts slope
occurs at 1 to 10 mJy, the brightness of typical $L_{*}$ galaxies
at redshifts greater than 1.
At longer wavelengths, between 200 and 400$\mu$m,
on the falling side of the peak of the rest-frame energy distribution,
there is hardly any flux range where
the counts slope is Euclidean.  The counts at the {\it bright}
end, from 1 to 0.01 Jy, actually rise more steeply than the
-1.5 slope, owing to the strong K-corrections for redshifts up to 
$\sim$ 1 to 2.
For long-wavelength fluxes below 0.01 Jy, the counts curves flatten
from the contributions of galaxies at redshifts of 2 and higher.

\subsection{Comparison with Other Models}

Our predicted counts at all wavelengths have a steeper slope
(more faint sources) than 
those of  Takeuchi {\it et al.} (1999), because our assumed evolution 
($Q = 3.1$) is twice as much as theirs ($Q = 1.4$ or 1.5).
We are in closer agreement with Takeuchi {\it et al.} (2001),
who used a stronger, but non-parametric luminosity evolution.
Their ``Evolution 1" scenario is similar to our baseline model at all
wavelengths,
while their ``Evolution 2" scenario is slightly closer to our Fast evolution model.
Our predicted baseline model counts agree with the 
(similar) ``baseline model" of Pearson \etal
(2001) at all wavelengths they considered (7 to 25\mic). 
At very faint flux levels around 0.1 mJy, our predictions agree with 
their modified model which included an {\it ad hoc} population of 
strongly evolving ``ultraluminous infrared galaxies".
However, our count predictions at the faintest fluxes (0.001 mJy)
are signinficantly above theirs, partly because of their adoption of
an open geometry for the Universe.
Our predicted 15, 60 and 175 $\mu$m counts are close to those of Tan 
{\it et al.} (1999),
although 10 to 40\% higher, because our assumed $Q = 3.1$
evolution is effectively stronger up to $z = 2$
than that of their disk galaxies (which dominate much of the counts at these 
wavelengths).
Our counts slopes are steeper than those of model E of Guiderdoni 
{\it et al.} (1998).
We predict two to three times more faint sources at 15$\mu$m, 
50\% more at 60$\mu$m, and only slightly more faint sources at 175$\mu$m.
The {\it shape} of our number count curves agrees very well with those of
Xu \etal (1998), as it must, because we both used virtually identical
methodology and luminosity evolution assumptions.
However, the {\it normalization} of our counts is higher by about a factor of
two (more at 15$\mu$m, and less at 25$\mu$m), because of the differences in
our assumed local LF's.

In Figures 5 and 6 we compare the predictions of our
three models with observations of faint integral source counts
at 15$\mu$m and 175$\mu$m, respectively.
The 15$\mu$m source count data are taken from the 
recent compilations by Serjeant \etal (2000) for the brighter fluxes
(shown by the thick line),
and Elbaz \etal (2000) for the fainter fluxes
and are mostly based on various deep {\it ISOCAM} fields.
The 175$\mu$m counts are taken from a compilation by Dole {\it et al.} (2000)
and Juvela {\it et al.} (2000),
based on deep {\it ISOPHOT} imaging in the {\it FIRBACK} (Far Infrared
Background) survey.
The three model predictions are again shown as a solid line for the baseline
case, a dashed line for the fast evolution case, and a dot-dashed line
for the lower limit case.
Our new description of the long-wavelength spectrum increased the predicted
number counts at 175 $\mu$m by about 30\% over our pre-{\it ISOPHOT} estimate. 
This change, of course, does not effect the counts at shorter wavelengths.

The observed faint counts at both mid- and far infrared wavelengths agree 
with our baseline model.
The only apparent discrepancy occurs around intermediate 15$\mu$m fluxes,
where the counts from the Marano fields fall significantly below comparable
measurements from other data sets.  It appears, in fact, that much of this
apparent disagreement is attributable to different choices of flux
calibration by the different {\it ISOCAM} teams (Elbaz {\it et al.} 2000).
The predictions of the
``Lower Limit" model with truncated evolution at $z_{flat}= 1$ is
so similar to the baseline model that they are also close to the
observations. With $Q = 3.1$, changing $z_{flat}$ from
2 to 1 decreases the predicted number of 0.1 mJy sources at 15$\mu$m 
by only 20\%.
With $z_{flat} = 1$, increasing $Q$ from 3.1 to 4.1 
results in a 30\% increase in bright (1 Jy) 175$\mu$m sources.
However, extending evolution with $Q = 4.1$ up to $z_{flat} = 1.3$  
leads to a prediction of 40 to 50\% more
faint sources, in apparent conflict with most of the observations.
Certainly a very fast evolution case ($Q = 5$) is ruled out, as it 
significantly overpredicts (by 75\% or more) 
the number of 15 and 175$\mu$m sources. 

Other recent studies have also used backwards-evolution models to predict
faint infrared source counts.
Pearson \etal (2000) and Xu \etal (2000) divided their galaxy populations
into a less luminous group having mild cosmic evolution,
and an "ultraluminous" galaxy class with stronger evolution than we assumed.
We have adopted the simpler assumption of a single evolution for all galaxies,
since current observations are too limited to provide serious constraints on
the additional free parameters that these models invoke.
The Xu \etal (2000) predictions agree with our ``fast evolution"
case for long wavelengths, but are significantly below at 6$\mu$m.
At wavelengths of 15$\mu$m and longer, the Xu \etal (2000) predictions 
rise a little more steeply than ours, due to their larger evolution index 
($Q = 4.5$), but then rise more slowly at the faintest fluxes due to their 
adoption of an open Universe geometry.

\section{Predicted Redshift Distributions for Future Surveys}

The range of plausible models calculated above all predict CIB spectra
and galaxy number counts at wavelengths from 3 to 400 $\mu$m 
which differ by at most a factor of 2.5. 
These models also differ in their
predictions of the redshift distributions of faint galaxies.
We now show how even limited redshift information for flux-limited
galaxy samples will discriminate between these models.

Several planned deep imaging surveys of low-foreground regions of the sky
should be sensitive enough to detect a significant fraction
of the galaxy population at moderate and very high redshifts.
In early 2003, the {\it SIRTF}  ``First Look Survey" 
should detect galaxies with fluxes of
1 mJy at 25 $\mu$m and 5 mJy at 70 $\mu$m with the Multiband Imaging 
Photometer, {\it MIPS}. 
{\it SIRTF} First Look Survey imaging with the Infrared Array Camera, 
{\it IRAC} 
will detect 30 $\mu$Jy sources at 6 $\mu$m.
In the same year, a wider sky area will be surveyed at 120$\mu$m by
{\it IRIS} (Shibai, 2000). In Figure 7 we assume that it will 
detect galaxies down to 32 mJy. Several years later, surveys with 
the  Photoconductor Array Camera and Spectrometer,    
{\it PACS} on the Far Infrared Space Telescope, {\it FIRST}
will detect 1 mJy sources at 200 $\mu$m.

The integrand of Eq. (4) evaluated at a fixed observing wavelength 
down to a given flux limit gives 
the differential redshift distribution, $ dN/dz$.
We plot this distribution in Figures 7a and 7b for sample deep observations
at 6, 25, 120 and 200 $\mu$m as a function of $z$.
The flux limits chosen at each wavelength are expected to be reached by planned
high galactic latitude imaging 
surveys with new space observatories in the next several years,
as discussed above.
The vertical axis is absolutely normalized to
one steradian of sky coverage. 
In Table 1, we list the actual numbers of galaxies per steradian
expected with redshifts above
1 and 2 in various flux-limited surveys.
The next two columns give the total number and number with $z \ge 1$
for the fast evolution scenario.

The counts at a given observed flux level are dominated
by the redshift at which $L_{*}$ galaxies appear.
As shown in Figure 3, the 120$\mu$m {\it IRIS} and 25$\mu$m {\it SIRTF}.
First Look 
surveys can both detect $L_{*}$ galaxies up to $z \sim 0.6$.
That is why the differential redshift distributions for
all galaxies detected in these two surveys are predicted to be similar,
as can be seen in Figure 7.
For the baseline evolution scenario, we predict that 18\% 
of the galaxies detected by {\it IRIS} will have $z > $1, and 2\% 
will have $z > $2. 
For the fast evolution scenario, 24\% of the detected galaxies 
will have $z >$1 and 1\% will have $z >$ 2.
{\it SOFIA} is expected to have comparable long-wavelength 
sensitivity to {\it IRIS} using 1 hour
integrations with chopping (Becklin 1997).

Assuming that the {\it MIPS} ``First Look" survey will cover five square 
degrees, our baseline model predicts integrated counts at 25 and 70$\mu$m 
of 3900 and 6100 galaxies, respectively. 
Down to those flux limits, the cosmological depths are comparable, 
with median redshifts of 
0.48 and 0.40, respectively, and upper quintile redshifts of 1.17 and 0.9.
The baseline model predicts 25\% and 5\% of galaxies at 25$\mu$m will have 
$z >$ 1 and $z >$ 2, respectively.  The fast evolution model
predicts these fractions are 33\% and 3\%.
Reaching a factor of 3 deeper at 25$\mu$m increases the depth to a median
redshift of 0.8 and an upper quintile redshift of 1.64.
 
In a five square degree First Look Survey at $6\mu$m with {\it IRAC}, there 
should be a total
of 9700 galaxies detected down to 30 $\mu$Jy.
Of these galaxies, 59\% should have redshifts greater
than 1, and 20 to 28\% should have redshifts greater than 2.
However, the reach would not be greatly extended by much longer time
integrations. Even at a limiting flux three times fainter (10 $\mu$Jy), the  
median redshift only increases from 1.30 to 1.44 and the upper
quintile redshift increases from $z = 2.3$ to 2.46.
This is because the First Look Survey with {\it IRAC} already reaches back to
$z_{flat}$.  Longer time integrations tend to detect more galaxies of lower
luminosity with a similar redshift distribution.

The sensitivity of {\it FIRST} is also sufficient to detect the
progenitors of modern $L_{*}$ galaxies at all redshifts out to 5 
(see Figure 3).
At wavelengths longward of 100$\mu$m,
the sensitivity to high-redshift galaxies 
is comparable to that expected for {\it IRAC} at 6 $\mu$m, owing to the 
positive K-corrections.
We have approximated the imaging sensitivity of the 
Spectral and Photometric Imaging Receiver {\it SPIRE} on 
{\it FIRST} as 3.2 mJy at
250 $\mu$m (Griffin 1997), which might also be 
approached for small areas of sky by {\it SOFIA}.
At this flux level, the strong positive K-correction
results in a remarkably far reach for detection of distant galaxies:
a median redshift of 1.62 and an upper quintile redshift of $z \ge 2.42$.
A {\it PACS} survey to 5 mJy at 150$\mu$m (Poglitsch 1997)
will have a median redshift of 0.93
and an upper quintile of $z \ge 1.70$, with
47\% of galaxies having $z >$ 1 and
12\% of galaxies having $z >$ 2.
It is gratifying to see that all of these planned surveys
should detect high surface densities of galaxies at these redshifts.

The predicted confusion noise limits for the beamsizes of each of these
instruments are also included in Table 1. 
These are very optimistically set to the flux levels at which there is an
average of one source per beam.  In reality, confusion becomes
a problem when there is one source per 10 to 20 beams, so that
the real confusion limits should probably be several times larger
than the numbers quoted in the Table.  The predictions are for the
baseline evolution; the confusion limit fluxes are slightly larger
in the fast evolution scenario. All of 
the planned surveys will be comfortably above the confusion limit
except {\it SPIRE} 250$\mu$m imaging, which is close to the confusion-limit,
and {\it IRIS} 120 and 150$\mu$m and {\it MIPS} 160$\mu$m imaging, 
which will be dominated by confusion noise.

\section{Conclusions}

We have used our empirically based model (MS98) 
to predict infrared luminosity
functions and deep infrared galaxy counts at various wavelengths.
We have also examined our predictions for the CIB for comparison with the
subsequent determinations from the {\it COBE-DIRBE} data analysis.
Using the formalism of luminosity evolution proportional to $(1+z)^Q$ out to
a redshift of $z_{flat}$ and constant (no further evolution) for 
$z_{flat} < z < z_{max} = 4$,
we find that a comparison of their predictions with current {\it ISO} 
galaxy counts at 15 and 175$\mu$m favor our ``Baseline Model" with $Q = 3.1$ 
and $z_{flat} = 2$ (the middle curve in Figure 2). The $\gamma$-ray 
limits (SD97) also favor $Q \sim 3$, as does a comparison of our predicted
CIB with the analysis of the {\it HEGRA} observations of 
high-energy $\gamma$-ray spectrum of Mrk 501 by 
Konopelko, Kirk, Stecker \& Mastichidas  (1999). 
On the other hand, the {\it COBE-DIRBE} 
far infrared determinations seem to favor a stronger evolution with 
$Q > 4$ up to $z_{flat} = 1$. For example, the upper curve in Figure 2
assumes $Q = 4.1$ and $z_{flat} = 1.3$.

This {\it prima facie} conflict can be resolved in two ways: either
(a) the {\it ISOPHOT} galaxy 
counts may be missing a significant fraction of sources. In this case,
one may also have to require that the $\gamma$-ray results are wrong in that
the energies of photons detected by
{\it HEGRA} have been overestimated, mimicing the effect which would be
caused by absorption from a lower CIB\footnote{Another
possibility is one involving new physics, {\it viz.} that Lorentz invariance
may be broken, allowing the Universe to be transparent to multi-TeV photons
(Coleman \& Glashow 1999; Kifune 1999; Glashow \& Stecker 2001; Stecker 2001).
This ``new physics''scenario presents 
problems in that the Mrk 501 spectrum does exhibit exactly the characteristics
expected for high-energy $\gamma$-ray absorption from pair-production 
(Konopelko {\it et al.} 1999).}, or
(b) the {\it COBE-DIRBE} far-infrared estimates may suffer from 
undersubtraction of foreground emission and therefore are too high.

If the far-infrared galaxy counts are incomplete (possibility (a)), 
this would imply stronger evolution in the far-infrared emission of 
galaxies than in the mid-infrared.
Although the MS98 model already includes some differential 
evolution of this type, based on the data of Spinoglio {\it et al.} (1995),
it is conceivable that starburst galaxies at redshifts $\sim 1$ might
produce an even higher ratio of $\sim 60\mu$m to $\sim 7\mu$m rest-frame 
fluxes than their present-day counterparts. 
Radiation at these widely separated wavelengths 
is known to be emitted by quite different dust grains which could have 
a different evolutionary development, particularly for ULIRGs (ultraluminous 
infrared galaxies) and AGN (active galactic nuclei). One should note that 
this possibility can make the $\gamma$-ray and infrared data compatible, 
since absorption of $\sim$ 15 TeV 
$\gamma$-rays is caused by interactions with mid-infrared ($\sim$ 20$\mu$m)
photons and not 140$\mu$m far-infrared photons.

Possibility (b) finds support in the independent analysis of the {\it COBE}
data by Lagache {\it et al.} (1999) who obtain a flux at 140 $\mu$m 
which is only 60\% of the flux obtained by 
Hauser {\it et al.} (1998) shown in Figure 1. 
Lagache {\it et al.} (2000) also obtained a smaller flux at 240 $\mu$m. 
In this regard, one should also note that the detections claimed by 
Hauser {\it et al.} (1998) were at the 4$\sigma$ level. 
Also, it is important to note that both our baseline model and our fast
evolution model are within 2 $\sigma$ of the {\it COBE-DIRBE} results obtained
by adopting the {\it FIRAS} calibrations of Fixsen, \etal (1997). In this case, 
{\it there would be no
inconsistency} with a $Q = 3.1$ evolution being acceptable to explain the
source counts, the CIB and the $\gamma$-ray mid-infrared upper limits.

We have shown here that planned infrared imaging surveys will soon be able
to measure galaxy evolution out to redshifts of 2 or greater and
help to resolve the question discussed above.
This can be done with our simple 
modelling technique in conjunction with 
observations of luminosity functions, galaxy counts, and the diffuse 
infrared background. 

A complete listing of the model count predictions at each wavelength 
can be obtained from the internet site  
{\email www.astro.ucla.edu/faculty/malkan.html}.

We would like to thank Michael Hauser, Ned Wright and Cong Xu for helpful 
discussions. We also thank an anonymous referee for helpful and constructive 
criticisms.

\begin{tabular}{|c|r|r|r|r|r|}
\multicolumn{6}{c}{\bf Table 1: Predicted Galaxy Counts With Future 
Instruments}\cr
\hline
Instrument ($\lambda(\mu$m)) & $F_{lim}^{a}$ & N(Tot)$^b$ & N($z > 1$) &
N($z > 2$) & Conf.Noise$^c$ 
\\ \hline 
MIPS/SIRTF (25) &  1.0 &  2.55(6) & 6.3(5) &  1.3(5) & 3.2 \\
" " &" & 4.57(6)$^d$ &1.5(5) & ... & \\
MIPS/SIRTF (25) & 0.32 & 1.34(7) & 5.5(6) & 1.5(6) & ... \\
MIPS/SIRTF (70) &  5.0 & 3.99(6)  & 6.7(5) & 1.1(5) &                130 \\
IRAC/SIRTF (6)  & .032 &  6.35(7) &  3.8(7) & 1.8(7) & 0.1 \\
" " & " & 8.86(7) &5.2(7) & ... & \\
IRAC/SIRTF (6) & .01 & 2.37(8) & 1.5(8) & 7.7(7) & ... \\
FIS/IRIS  (120) & 32 & 1.13(6) & 2.1(5) & 2.6(4) & 21000 \\
" " & " & 2.17(6) &7.5(5) & ... & \\
FIS/IRIS  (150) & 50 & 7.38(5) & 2.0(5) & 3.0(4) & ... \\
PACS/First (100) & 5.0 & 9.94(6) & 2.8(6) & 4.8(5) &                 40 \\  
PACS/First (150) & 5.0 & 2.16(7) & 1.0(7) & 2.5(6) &               400 \\
SPIRE/First (250) & 3.2 & 5.45(7) & 4.2(7) & 1.8(7) &               1100 \\
\hline
\end{tabular}

\medskip

\noindent (a) Expected limiting fluxes are for 5$\sigma$-detections, in units of mJy.

\noindent (b) Counts are given in scientific notation, with the exponent of 10 given in parentheses.

\noindent (c) Confusion Limits, given in $\mu$Jy, at which there is an average of one source
per beam (FWHM).

\noindent (d) The second line for each survey shows the galaxy number count preditions for the 
Fast Evolution scenario. At redshifts above $z = z_{flat} = 1.3$, there is no difference from the
baseline model.
             
\begin{figure}
\epsfxsize=16cm
\epsfbox{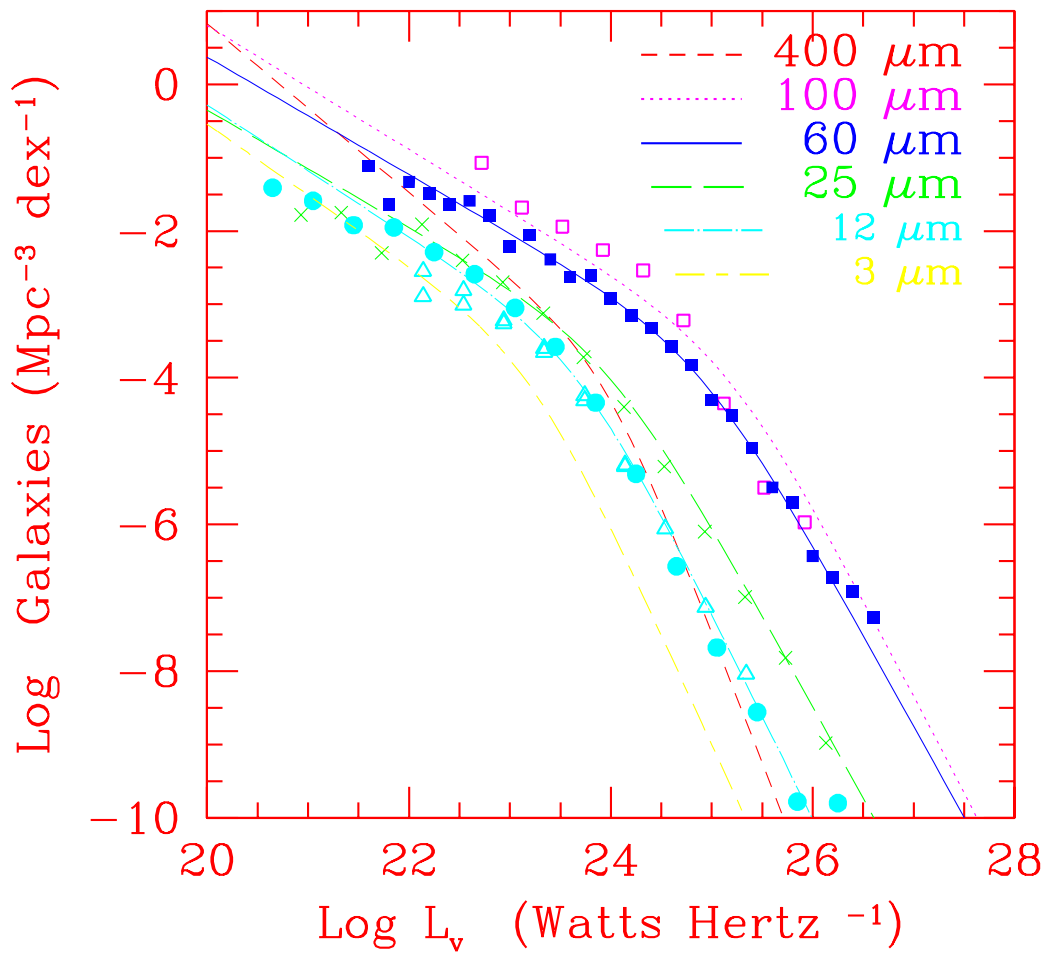}
\caption{The family of infrared luminosity functions generated 
as described in the text. These are broken power-laws, where the characteristic
turnover ``knee" varies with wavelength. The solid squares show observations at
60$\mu$m of Lawrence, \etal (1986), to be compared with the solid line.
The open boxes are from Soifer and Neugebauer (1991), for 100$\mu$m. 
The $\times$'s show data from 25$\mu$m of Shupe \etal (1998), to be compared
with the long-dashed line given by the model. The solid circles and open 
triangles show observations of the 12$\mu$m luminosity
function from Rush \etal (1993) and Fang \etal (1998) respectively, to be
compared with the dot--dash line.  Two other model LF's are plotted for
3 and 400$\mu$m,
wavelengths at which there is no current observational comparison available.}
\end{figure}

\begin{figure}
\epsfxsize=16cm
\epsfbox{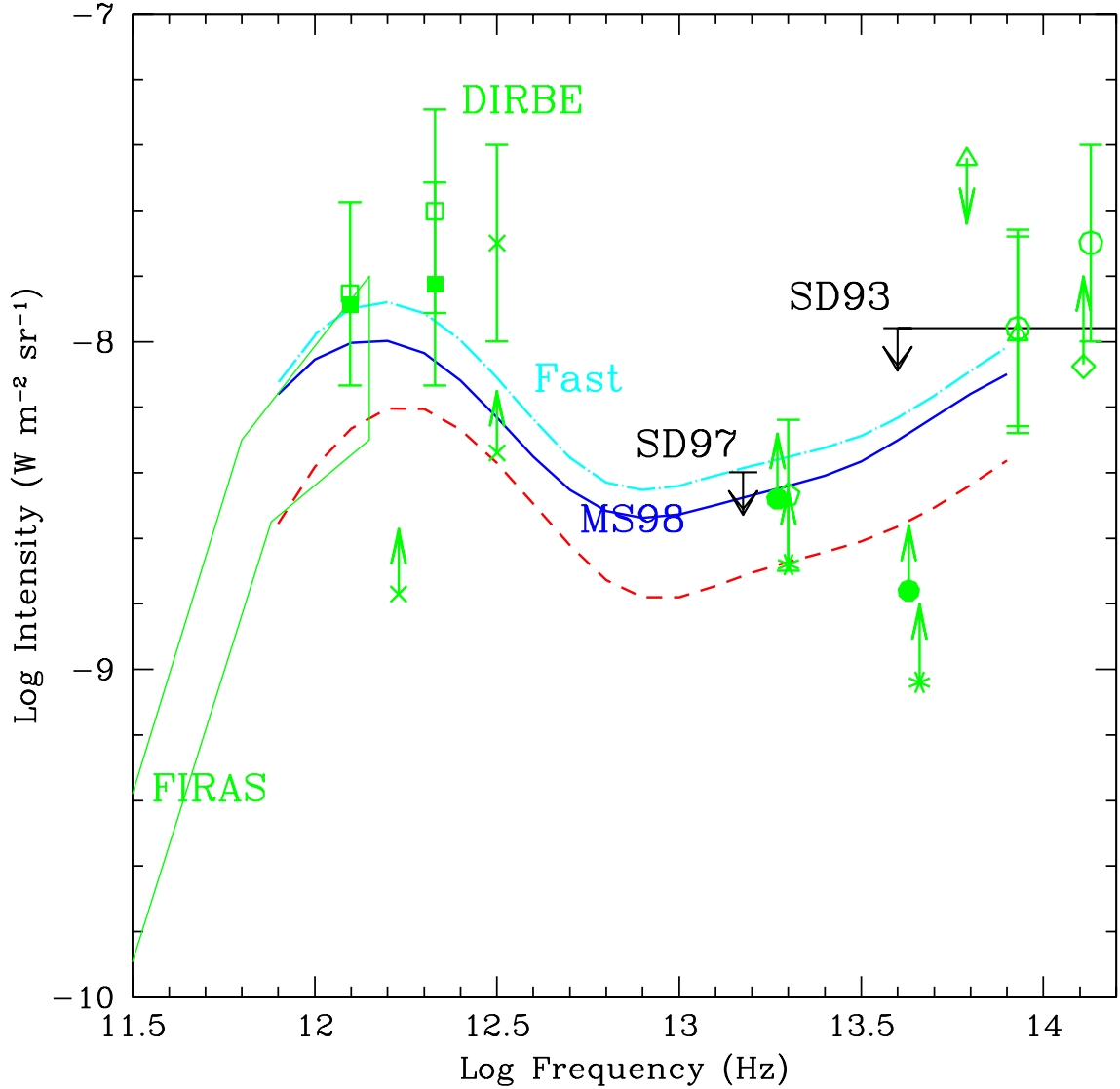}
\caption{SEDs of the 
CIB calculated using the models described in the text 
and compared with representative data given with 2$\sigma$ error bars.
The FIRAS range is illustrated with a polygon. 
>From left to right, open squares: Hauser \etal (1998), solid squares:
DIRBE data with FIRAS flux calibration, $\times$: 
Lagache (2000), solid circles: Altieri \etal (1999), open triangles:
Dwek and Arendt (1998), open circles: Gorjian, Wright \& Chary (2000), 
diamond: Totani, \etal (2001) (offset for clarity;
see also Pozzetti and Madau (2001). 
The lower limits with $\times$'s  are based on integrals of deep source counts
from {\it ISOPHOT} (Puget \etal 1999) 
and the lower limits with asterisks are from ISOCAM (Elbaz \etal (1999).
The upper limits are from Stecker \& de Jager (1993)(SD93) and Stecker \& 
de Jager (1997)(SD97).}
\end{figure}
             
\begin{figure}
\epsfxsize=16cm
\epsfbox{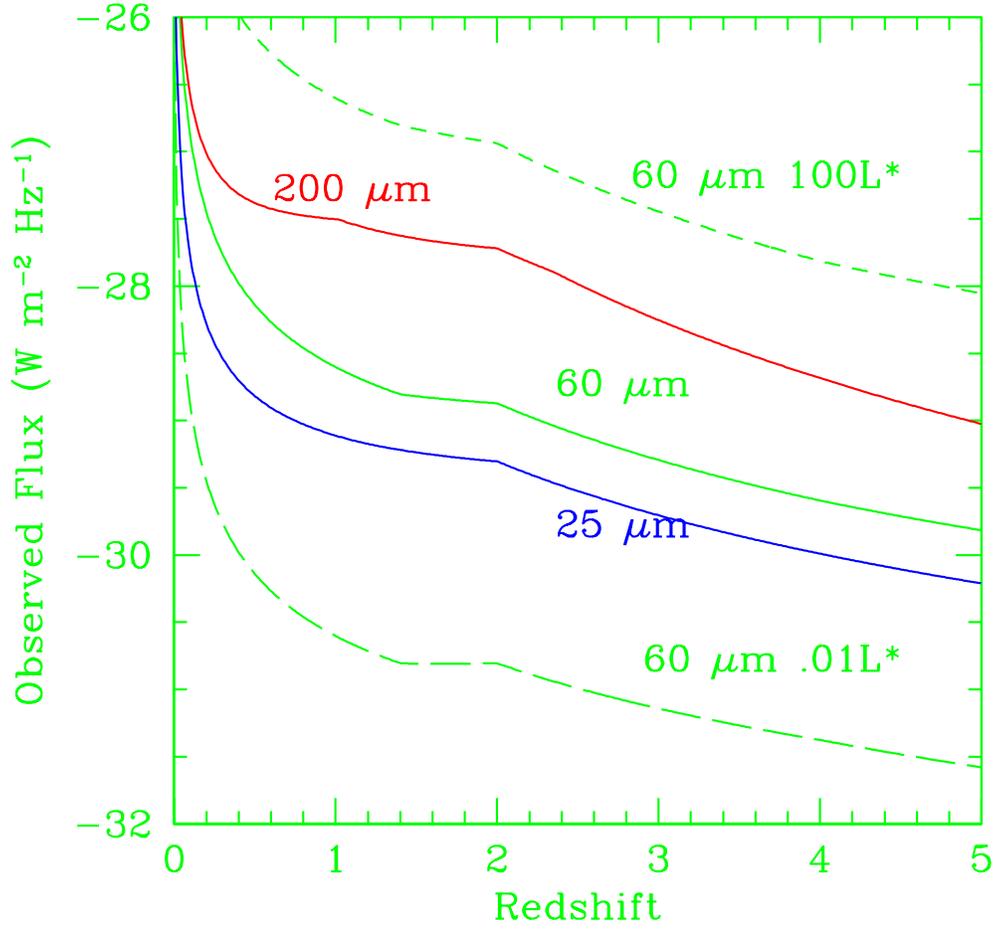}
\caption{The observed flux as a function of redshift for 
galaxies with $L_{*}$ at $z=0$ 
at 25, 60 and 200 $\mu$m, and at 60 $\mu$m for galaxies of 
luminosities 0.01 $L_{*}$ and 100 $L_{*}$ (dahsed lines). These estimates 
are for the
baseline evolution of $Q=3.1$ up to $z_{flat}=2$.}
\end{figure}

\begin{figure}
\epsfxsize=16cm
\epsfbox{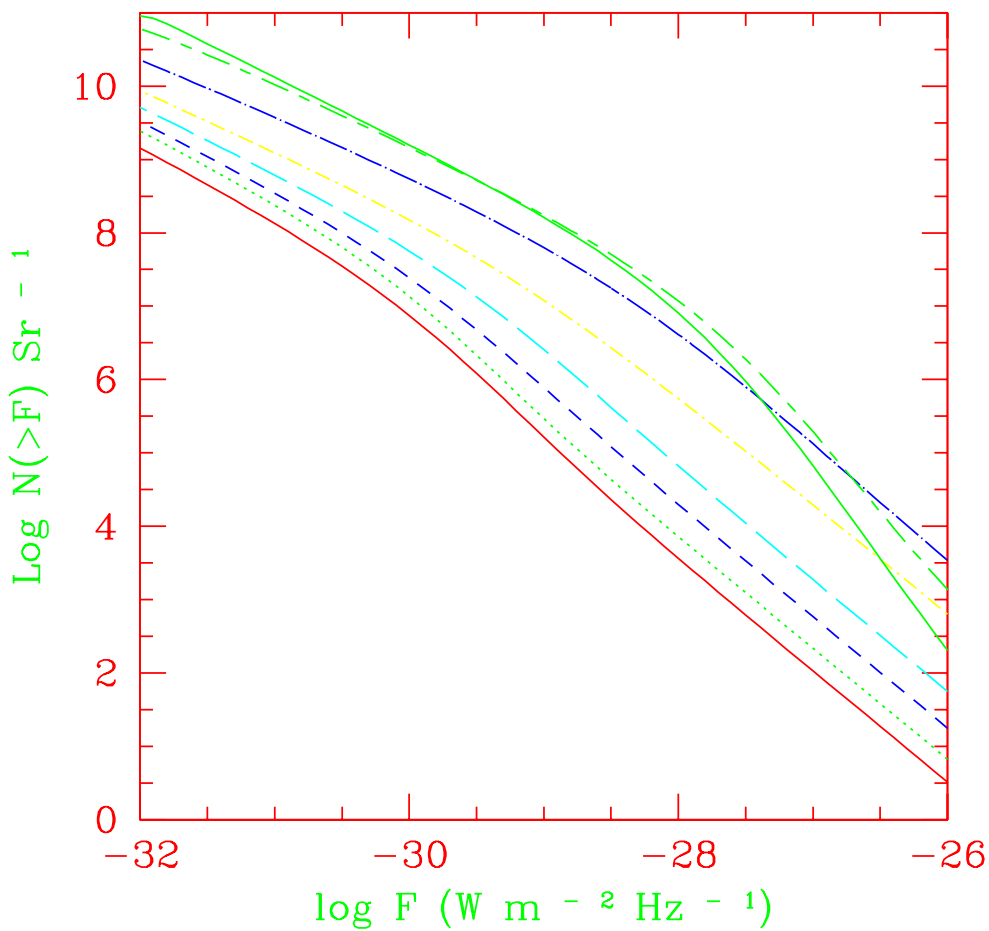}
\caption{Galaxy count predictions for the baseline model at  eight 
representative 
wavelengths. Solid lines are at 3 and 400$\mu$m (bottom and top).
The dotted line is for observations at 6$\mu$m; short dashes 12$\mu$m;
long dashes 25$\mu$m, dot-dash 50$\mu$m, dot-long dash 100$\mu$m,
and long dash--short dash 200$\mu$m. Except at the longest wavelengths,
most of the models show the expected Euclidean slope of -3/2 at the
bright end. All models turn flatter than this at the faintest fluxes
due to cosmological factors.}
\end{figure}
            
\begin{figure}
\epsfxsize=16cm
\epsfbox{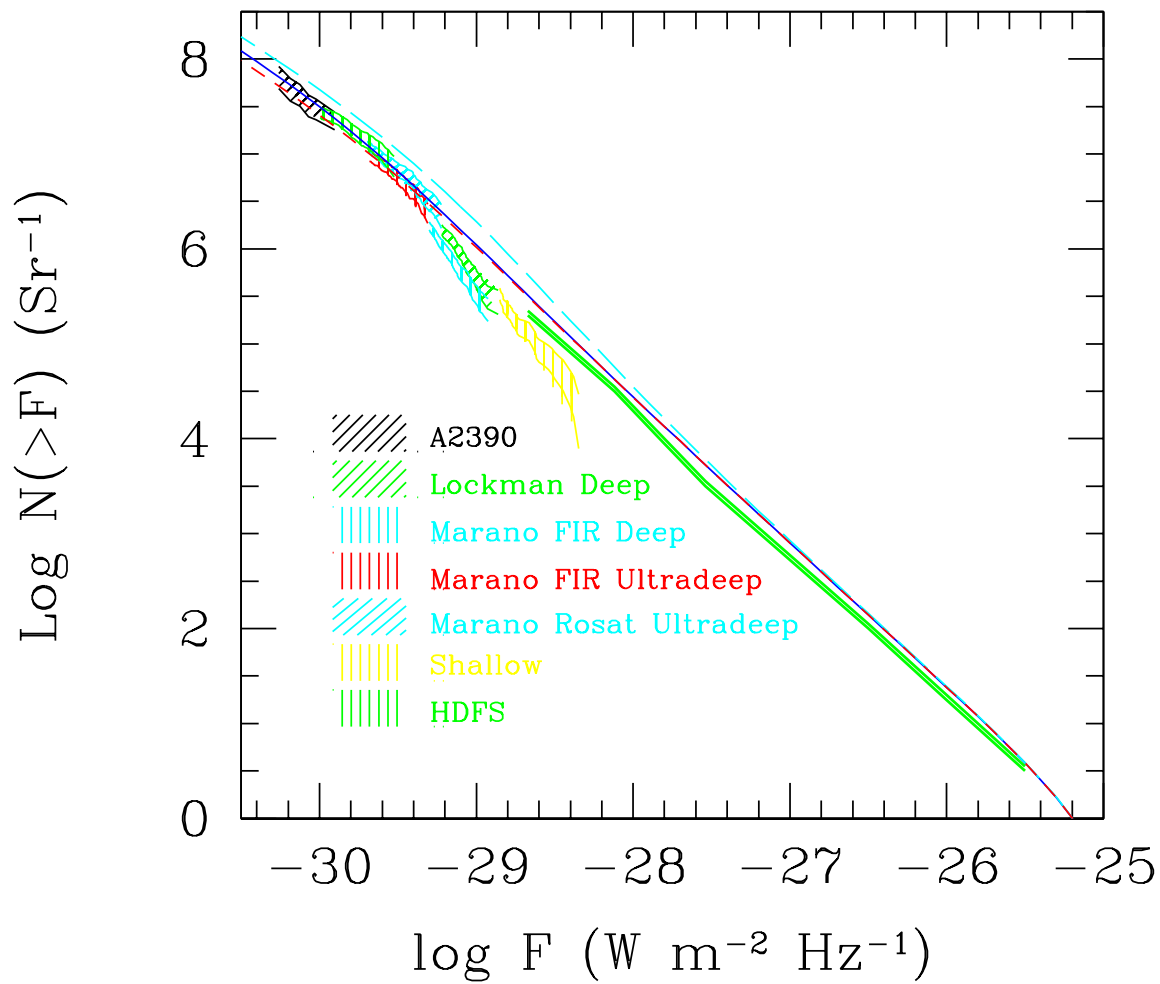}
\caption{Galaxy counts predicted for our various evolution models
at 15 $\mu$m compared with the data compiled by Elbaz \etal (2000),
and from Sejeant \etal (2000: thick line).
Error ranges of 1$\sigma$ are shown.  The solid line shows the baseline model;
the dashed line is the lower limit ($z_{flat}=1$), while the dot-dash
line shows the fast evolution model ($Q=4.1$ up to $z_{flat}=1.3$).}
\end{figure}
            
\begin{figure}
\epsfxsize=16cm
\epsfbox{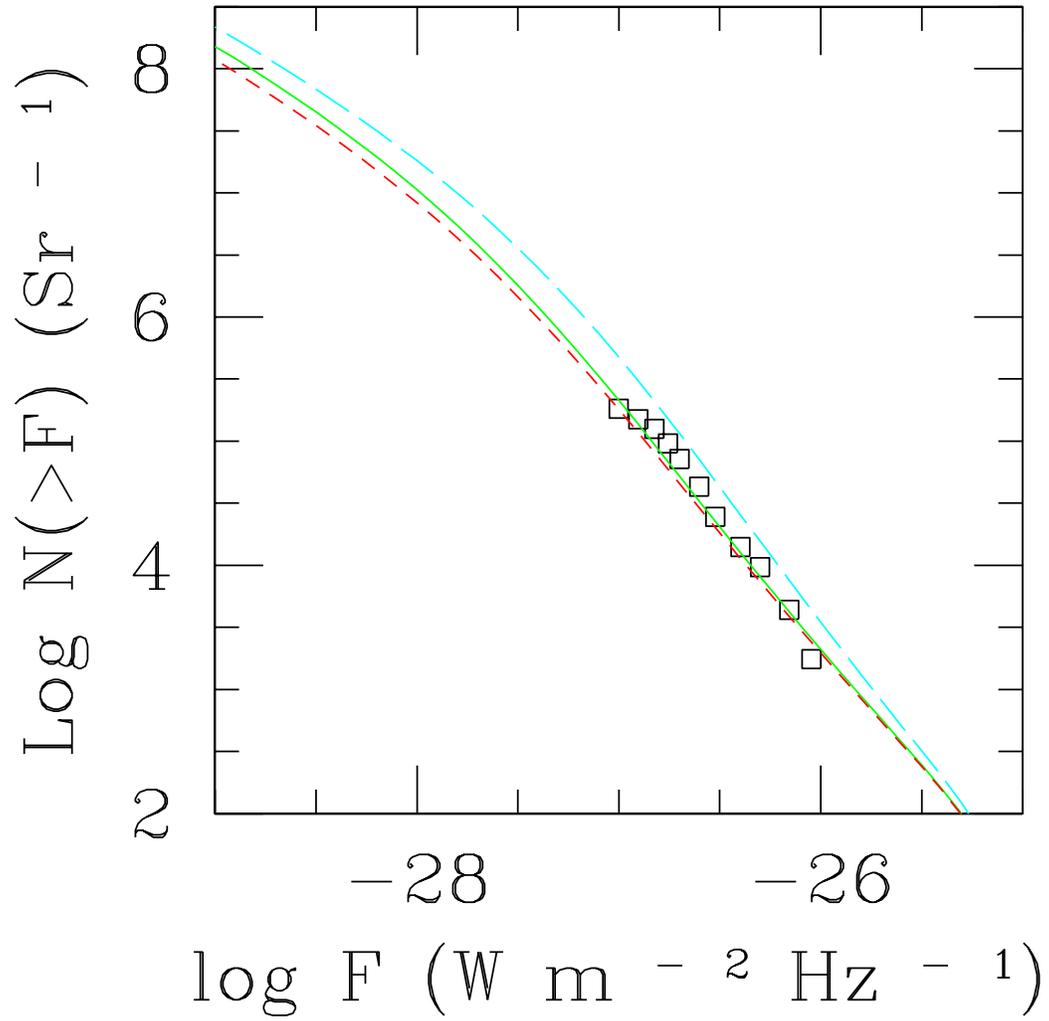}
\caption{Galaxy counts at 175 $\mu$m predicted for our various 
evolution models compared with the data (open squares) 
compiled by Puget \etal (1999). Models are illustrated with the same line
styles as in Figure 5.}
\end{figure}

\begin{figure}
\epsfxsize=16cm
\epsfbox{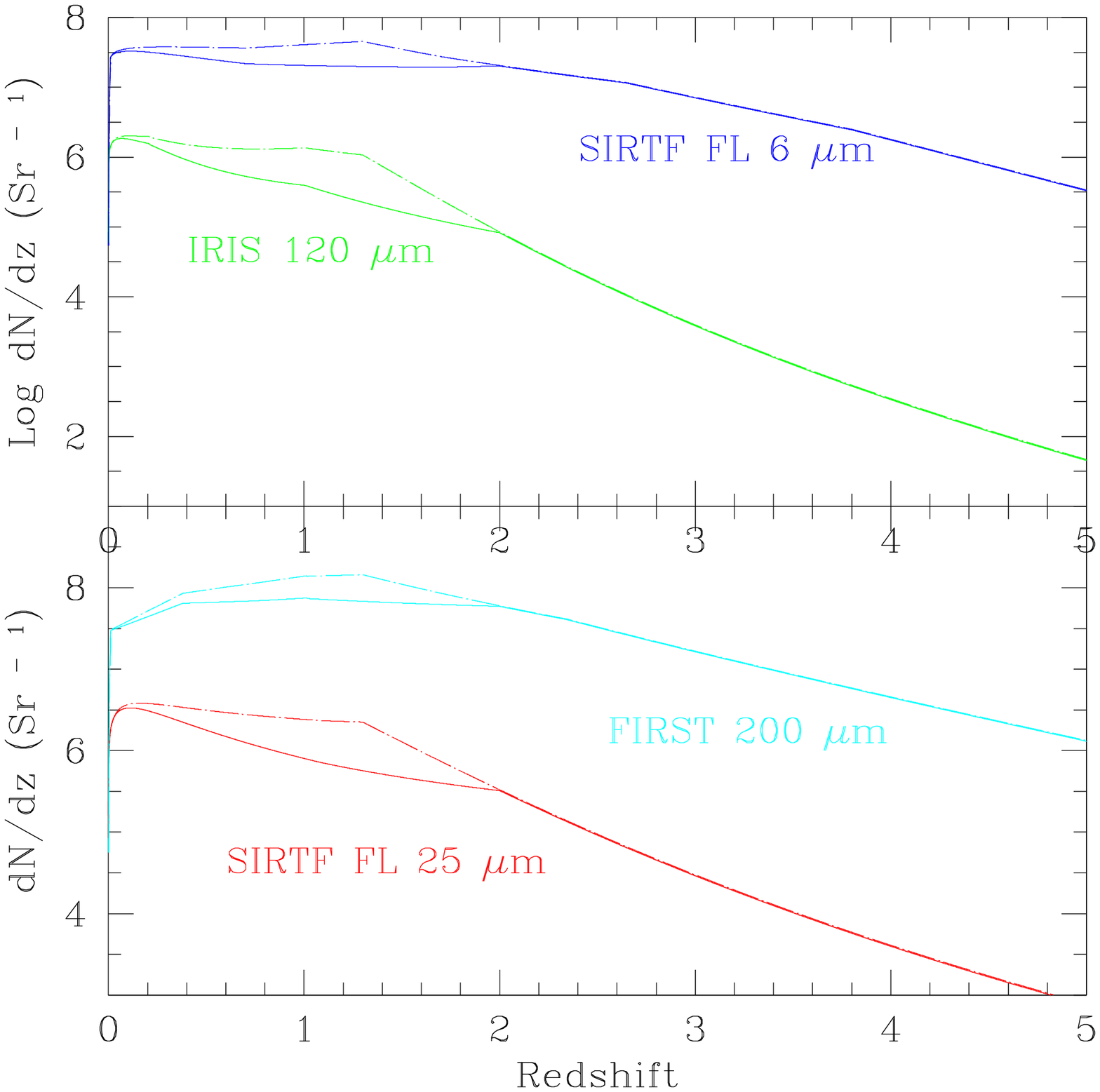}
\caption{Number of observable sources in flux-limited samples
covering one Steradian of sky, 
as a function of redshift 
for planned space infrared telescopes as described in the text. The solid lines
are for the baseline model. The dot-dash line are for the fast evolution case,
which has the greatest excess of galaxies at its value of $z = z_{flat} = 1.3$.
Note that the {\it SIRTF} First Look Survey at 6$\mu$m and 
{\it FIRST} survey at
200$\mu$m have comparable depths, as do the {\it IRIS} and {\it SIRTF/MIPS} 
surveys.}
\end{figure}


\newpage



\begin{thebibliography}{}

\bibitem{al99} Altieri, B. \etal 1999, A \& A 343, L65

\bibitem{be97} Becklin, E.E.  1997, Proceedings of ESA Symposium on 
``The Far-Infrared and Submillimeter Universe", pg. 201

\bibitem{bi98} Biller, S.D. \etal 1998, Phys. Rev. Lett. 80, 2992

\bibitem{bl00} Blain, A.W. \& Natarajan, P. 2000, MNRAS, 312, L39

\bibitem{co99} Coleman, S. \& Glashow, S.L. 1999, Phys. Rev. D59, 116008

\bibitem{do00} Dole, H \etal 2000, astro-ph/0002283

\bibitem{dw98} Dwek, E. \& Arendt, R. 1998, ApJ 508, L9

\bibitem{el00} Elbaz, D. \etal 2000, A \& A Lett., in press, e-print 
astro-ph/9910496

\bibitem{fa98} Fang, F., Shupe, D.L., Xu, C. \& Hacking, P.B. 1998, ApJ 500, 693

\bibitem{fi00} Finkbeiner, D.P., Davis, M. \& Schlegel, D.J. 2000, ApJ 544, 81

\bibitem{fi01} Finkbeiner, D.P. 2001, in {\it The Extragalactic Background
and its Cosmological Implications}, IAU Symp. Vol. 204, ed. M. Harwit \& M.G.
Hauser, in press.

\bibitem{fi97} Fixsen, D.J. \etal 1997, ApJ 490, 482

\bibitem{fi98} Fixsen, D.J. \etal 1998, ApJ 508, 123

\bibitem{fr98} Franceschini, A., \etal 1998, MNRAS 296, 709

\bibitem{gi00} Gispert, R, Lagache, G. and Puget, J. 2000, A \& A, 360, 1

\bibitem{gl00} Glashow, S.L. \& Stecker, F.W. 2001, in preparation

\bibitem{go00} Gorjian, V., Wright, E.L. \& Chary, R.R. 2000, ApJ, 536, 550

\bibitem{gr97} Griffin, M. J. 1997, presentation at Grenoble meeting April 1997,
{\email http://astro.estec.esa.nl/First}

\bibitem{gu98} Guiderdoni, B., Hivon, E., Bouchet, F. \& Maffei, B. 1998, 
MNRAS 295, 877

\bibitem{ha98} Hauser, M., \etal 1998, ApJ 508, 25

\bibitem{ha01} Hauser, M. \& Dwek, E. 2001, {\it Ann. Rev. Astron. Ap.}, in press

\bibitem{ho00} Hopkins, A.M., Connelly, A.J. \& Szalay, A.S. 2000, 
AJ, in press, e-print astro-ph/0009073 

\bibitem{ju00} Juvela, M., Mattila, K. and Lemke, D. 2000, A \& A 360, 813

\bibitem{ki99} Kifune, T. 1999, ApJ 518, L21

\bibitem{ko99} Konopelko, A.K., Kirk, J.G. Stecker, F.W. \& Mastichiadas, A. 1999,
ApJ 518, L13  

\bibitem{la00} Lagache, G., Haffner, L.M., Reynolds, R.J. \& Tufte, S.L.
2000, A \& A 354, 247

\bibitem{la86} Lawrence, A. \etal 1986, MNRAS 219, 687

\bibitem{ma98} Madau, P., Pozzetti, L. \& Dickinson, M. 1998, ApJ 498, 106

\bibitem{ma99} Malkan, M.A., 1998 in {\it Astrophysics with Infrared 
Arrays: A Prelude to SIRTF}, Astron. Soc. Pacific, Conf. Series, in press,
e-print astro-ph/9810055

\bibitem{ma00} Malkan, M.A., 2000, 
in {\it Our Second Look at the Immature Universe: 
The Infrared View}, Proc. Fourth RESCEU International Symposium, 
(Tokyo: Universal Academy Press) in press,
e-print, astro-ph/0005251

\bibitem{ms98} Malkan, M.A. \& Stecker, F.W. 1998, ApJ 496, 13 (MS98)

\bibitem{pe01} Pearson, C.P. \etal 2001, astro-ph/0008472v3, submitted to
MNRAS

\bibitem{po97} Poglitsch, A. 1997, presentation at Grenoble meeting April 1997,
{\email http://astro.estec.esa.nl/First}

\bibitem{po01} Pozzetti, L. \& Madau, P. 2001,  in {\it The Extragalactic 
Background and its Cosmological Implications}, IAU Symp. Vol. 204, ed. 
M. Harwit and M.G. Hauser, in press, astro-ph/0011359

\bibitem{pu99} Puget, J.L. \etal 1999, A \& A 354, 29

\bibitem{ru93} Rush, B., Malkan, M.A. \& Spinoglio, L. 1993, ApJS 89, 1

\bibitem{sa90} Saunders, W., \etal 1990, MNRAS 242, 318 

\bibitem{se00} Serjeant, S. \etal 2000, MNRAS, 316, 768

\bibitem{sh00} Shibai, H. 2000, Adv Space Res, 25, 2273

\bibitem{sh98} Shupe, D., Fang, F., Hacking, P.B.\& Huchra, J.P. 1998, ApJ 501, 
597  

\bibitem{so87} Soifer, B.T., \etal 1987, ApJ 320, 238

\bibitem{so91} Soifer, B.T. and Neugebauer, G. 1991, AJ, 101, 354

\bibitem{sp95} Spinoglio, L., \etal 1995, ApJ 453, 616

\bibitem{sp01} Spinoglio, L., \etal 2001,  in preparation

\bibitem{st98} Stanev, T. \& Franceschini, A. 1998, ApJ 494, 159

\bibitem{st01} Stecker, F.W. 2001, in {\it The Extragalactic Background and 
its Cosmological Implications}, IAU Symp. Vol. 204, ed. M. Harwit and M.G. 
Hauser, in press, e-print astro-ph/0010015

\bibitem{st93} Stecker, F.W. \& De Jager, O.C. 1993, ApJ 415, L71 

\bibitem{st97} Stecker, F.W. \& De Jager, O.C. 1997, in {\it Towards a
Major Atmospheric Cerenkov Detector V}, Proc. Kruger National Park
Workshop on TeV Gamma Ray Astrophysics, ed. O.C. De Jager (Potchefstroom:
Wesprint) pg. 39

\bibitem{st99} Steidel, C.C. \etal 1999, ApJ 519, 1

\bibitem{ta99} Takeuchi, T.T. \etal 1999, PASP 111, 288

\bibitem{ta01} Takeuchi, T.T. \etal 2001, astro-ph/0009460, Pub. Astron. Soc. Japan,
in press

\bibitem{ta00} Tan, C., Silk, J. and Balland, C. 2000, ApJ 522, 579

\bibitem{to01} Totani, T. \etal 2001, in {\it The Extragalactic Background and 
its Cosmological Implications}, IAU Symp. Vol. 204, ed. M. Harwit and M.G. 
Hauser, in press

\bibitem{xu98} Xu, C. \etal 1998, ApJ 508, 576 

\bibitem{xu01} Xu, C. \etal 2001, astro-ph/0009220, ApJ, in press

\end{thebibliography}
\end{document}